\documentclass[12pt]{amsart}
 
\usepackage{amsmath}
\usepackage{color}
 \usepackage{epsfig}
\usepackage{amssymb}
\usepackage{amsfonts}
 \usepackage{latexsym}
\usepackage{subeqnarray}

    \def\RR{\mathbb{R}}

\begin{document}

%\centerline {DRAFT, costin/stillinger/040904.tex}

%\noindent Corrected by JLL 4/12/04 afternoon on fromovidiu031604.tex

\title{On the Construction of Particle Distributions with Specified Single
and Pair Densities}
\author{O. Costin, and J. L. 
 Lebowitz$^1$
{$_{\mbox{ Department of Mathematics, Rutgers University}}$}}
\thanks{$^1$Also Department of Physics.}
\gdef\shorttitle{Point Processes with Specified Density}
\gdef\shortauthors{O. Costin and J. L. Lebowitz}

\maketitle

\begin{abstract}

We discuss necessary conditions for the existence of probability
distribution on particle configurations in $d$-dimensions i.e.\ a
point process, compatible with a specified density $\rho$ and radial
distribution function $g({\bf r})$.  In $d=1$ we give necessary and
sufficient criteria on $\rho g({\bf r})$ for the existence of such a
point process of renewal (Markov) type.  We prove that these
conditions are satisfied for the case $g(r) = 0, r < D$ and $g(r) = 1,
r > D$, if and only if $\rho D \leq e^{-1}$: the maximum density
obtainable from diluting a Poisson process.  We then describe briefly
necessary and sufficient conditions, valid in every dimension, for
$\rho g(r)$ to specify a determinantal point process for which all
$n$-particle densities, $\rho_n({\bf r}_1, ..., {\bf r}_n)$, are given
explicitly as determinants.  We give several examples.
\end{abstract}

\section{Introduction}

The microscopic structure of macroscopic systems, such as fluids, is
best described by the joint $n$-particle densities $\rho_n({\bf r}_1,
... {\bf r}_n)$ where the ${\bf r}_1,...,{\bf r}_n$ are position
vectors in $d$-dimensions \cite{Hansen-McDonald}.  The most important
of these
 are the
one particle density $\rho_1({\bf r}_1)$ and the pair density
$\rho_2({\bf r}_1,{\bf r}_2)$.  {}For spatially homogeneous systems, the
only ones we shall consider here, {}From these one can
also obtain an infinite set of conditions in the case where only
$\rho_1$ and $\rho_2$ are given \cite{Kuna}.  These conditions are
very hard (or impossible) to check so the real question is whether one
can get away with a smaller number of readily checkable conditions.

A simple subset of such positivity conditions, emphasized by Percus
\cite{Percus} and by Stillinger, Torquato, et
al. \cite{Stillinger-Torquato-Eroles}-\cite{Stillinger-Torquato},
which follow directly from the definitions are,
%\begin{eqnarray}
\begin{equation}
\rho > 0, \quad g(\bf r) \geq 0,
\label{1.1}\
\end{equation}
\begin{equation}
\hat S({\bf k}) = \rho +
\rho^2 \int_{{\RR}^d} e^{i{\bf k} \cdot {\bf r}}[g({\bf r})-1] d{\bf r} \geq 0
\label{1.2}  
%\end{eqnarray}
\end{equation}

Conditions (\ref{1.1}) are obvious while (\ref{1.2}) ensures that
variances of one particle sum functions, $ \psi = \sum \phi({\bf
r}_i)$, are non-negative, since it follows from the definition of
$\rho_2$ that 
\begin{equation}
\langle (\psi - \langle \psi \rangle)^2\rangle = \Big( \frac{1}{2\pi}
\Big ) ^d \int d{\bf k} | \hat \phi({\bf k})|^2 \hat S({\bf k})
\label{1.3}
\end{equation}
where $\hat \phi({\bf k}) = \int e^{i{\bf k}\cdot{\bf r}} \phi({\bf r})d{\bf
r}$, and the averages are with respect to the probability distribution
of the point process.  Simple considerations, see \cite{Yamada}, show that one
should add to these (\ref{1.1}) and (\ref{1.2}) at least one further
requirement: the variance $V_\Lambda$ of the number of 
particles $N_\Lambda$ in a region $\Lambda$, which corresponds to 
$\phi({\bf r})$ being the characteristic function of the region
$\Lambda$, must be larger than 
$\theta(1 - \theta)$, where $\theta$ is the fractional part of the
average number of particles in $\Lambda$.  That is, if $\langle
N_\Lambda \rangle = k + \theta$, for $k$ a non-negative integer,
$k=0,1,2,...$, then
\begin{equation}
V_\Lambda = \langle (N_\Lambda - 
\langle N_\Lambda \rangle)^2 \rangle \geq \theta(1 - \theta).
\label{1.4}
\end{equation}
The bound (\ref{1.4}) comes from the fact that $N_\Lambda$ can only
take non-negative integer values, see \cite{Yamada} and Appendix where
a more general condition of type (\ref{1.4}) is proven.

A simple one dimensional example for which (\ref{1.1}) and (\ref{1.2})
are satisfied but (\ref{1.4}) is violated, is
\begin{equation}
g(r) = \left\{\begin{array}{ll} \displaystyle
   2(\rho r)^3, &\mbox {$r < \rho^{-1}$}
      \\
  \displaystyle 
   1, &\mbox{$r > \rho^{-1}$}
\label{1.5}
\end{array}\right.\end{equation}
A direct computation shows that $\hat S(k) \geq 0$, with equality
holding for $k=0$, (see below), while the variance in an interval of
length $L$, $L > \rho^{-1}$ is equal to $1/5$, which violates
(\ref{1.4}), whenever $\theta(1 - \theta) > 1/5$.  It is possible that
the condition (\ref{1.4}) becomes less important in higher dimensions
where the minimal variance will go to infinity as the domain grows.
{}For spherical domains it will grow at least like the surface area
\cite{Beck}, \cite{Lebowitz}.  Note that by choosing the value of $r$
beyond which $g(r) = 1$ as slightly smaller than $\rho^{-1}$
(\ref{1.1}) and (\ref{1.2}) would be satisfied but (\ref{1.4}) would
not for some $L$.

It is however possible, especially for the type of $g(r)$ considered
in \cite{Stillinger-Torquato-Eroles}--\cite{Stillinger-Torquato} that
(\ref{1.1}) and (\ref{1.2}) are enough to ensure realizability.  These
$g(r)$ have a hard core exclusion, prohibiting the centers of two
particles from coming closer than a certain distance $D$, i.e.\ $g(r)
= 0$ for $r < D$.  In particlar, it was conjectured in
\cite{Stillinger-Torquato-Eroles}--\cite{Torquato-Stillinger2003} that
it is possible to find a point process with density $\rho > 0$ and a
$g(r)$ of the form
\begin{equation}
g(r) = \left\{\begin{array}{ll} \displaystyle
   0, &\mbox {$r < D$}
      \\
  \displaystyle 
   1, &\mbox{$r > D$}
\label{1.6}
\end{array}\right.\end{equation}
as long as $\rho v(D) \leq 2^{-d}$, where $v(D)$ is the volume of a
$d$-dimensional sphere of radius $D/2$.  {}For $\rho v(D) > 2^{-d}$,
$\hat S({\bf k})$ will be negative for ${\bf k} = 0$.  (\ref{1.6})
also satisfies condition (\ref{1.4}) for $\rho v(D) \leq 2^{-d}$,
although this was not explicitly imposed.  There are also heuristic
arguments, bolstered by computer simulations
\cite{Crawford-Torquato-Stillnger} and by considerations of the $d \to
\infty$ limit \cite{Torquato-Stillinger2004}, for the realizability of
(\ref{1.6}) when $\rho v(D) \leq 2^{-d}$.

The case $\rho v(D) = 2^{-d}$, for which $\hat S(0) = 0$, is of
particular interest, since it yields a system for which the variance
$V_\Lambda$ grows only like the surface area of the boundary of
$\Lambda$ rather than the volume.  Such systems are variously called
superhomogeneous \cite{Goldstein-Lebowitz-Speer},
\cite{Aizenman-Goldstein-Lebowitz}, or hyperuniform
\cite{Torquato-Stillinger2003}.  In $d=1$, the variance in an interval
of length $L$, $V_L$ can actually be bounded uniformly in $L$ as
analyzed in \cite{Goldstein-Lebowitz-Speer},
\cite{Aizenman-Goldstein-Lebowitz} . Thus for the example (\ref{1.6})
with $\rho D = 1/2$,
\begin{equation}
V_L = \left\{\begin{array}{ll} \displaystyle
\rho L(1 - \rho L), & \mbox{$\rho L \leq \frac{1}{2}$}
      \\
  \displaystyle 
\frac{1}{4}, &\mbox{$\rho L \geq \frac{1}{2}$}
\end{array}\right.
\label{1.7}
\end{equation}
This is, by (\ref{1.4}), the minimum permissible variance when $\rho L 
= k + \frac{1}{2}, k = 0,1,2,...$ .

Inspired by the work of Stillinger and Torquato we give here a proof
of realizability of the model $g(r)$ in (\ref{1.6}) for the case $d=1$
and $\rho v(D) = \rho D \leq e^{-1}$.  This is based on a particular
construction of the point process as a dilution of a Poisson process
with $\rho D = \lambda D \exp[-\lambda D]$.  It turns out that the new
process is a Markov or renewal process \cite{Daley}--\cite{Cox-Isham}.
This permits us to describe all higher order correlation functions in
terms of $g(r)$.  We do not know at present whether there exist non
renewal point processes for some or all $\rho D \in (e^{-1},
\frac{1}{2}]$.  We also do not know whether the explicitly constructed
process for $\rho D \leq e^{-1}$ is unique.  In principle there can
exist more than one process with the same $\rho$ and $g(r)$ but
different higher order correlations; see sec.\ 5.

We note that one dimensional renewal processes, described in sec.\ 2,
and determinantal processes for arbitrary dimension, described in
sec.\ 4, are the only examples we know for which one can explicitly
(and easily) construct higher order correlations from $\rho_1$ and
$\rho_2$.  In some cases these processes correspond to the
distribution of particles in equilibrium systems.  There is also a 
formula for the entropy of a renewal process in terms of $g(r)$
\cite{Cox-Isham}.  

\section{Renewal Processes}
\setcounter{equation}{0} A translational invariant one dimensional
particle system with density $\rho > 0$, is described by a renewal
process (RP) whenever the conditional probability density of finding a
particle (or point) at a position $q$ on the line, given the
configuration of all particles to the left of $q$, say, $..., q_{-1} <
q_0 < q$, depends only on $x = q - q_1$ \cite{Daley},
\cite{Cox-Isham}.  Let us call that density $P_1(x)$.  In other words,
given that there is a particle at $q$, $P_1(x)$ is the probability
density that the {\it first} particle to the right (left) of $q$ is at
$q+x$ ($q-x$).  This corresponds, if we think of the points as events
in time, to a Markov process.  Clearly

\begin{equation}
\int_0^\infty P_1(x) dx = 1, \quad \int_0^\infty x P_{~~}(x) dx =
\rho^{-1}
\label{2.1}
\end{equation}

Calling $P_n(x)$ the probability density for finding the $n$th
particle at a distance $x$ to the right of the specified position of a
given particle 
we have
\begin{equation}
P_n(x) = \int_0^x P_{n-1}(x-y) P_1(y) dy, \quad n=2,3,...
\label{2.2a}
\end{equation}
By the definition of $\rho g(r)$ we have
\begin{equation}
\rho g(r) = \sum_{n=1}^\infty P_n(r)
\label{2.2}
\end{equation}
Taking the Laplace transform of (\ref{2.2}), using (\ref{2.2a}), then
gives
$$
\rho \bar g(s) \equiv \rho \int_0^\infty e^{-sr} g(r) dr =
\sum_{n=1}^\infty [\bar P_1(s)]^n
$$
\begin{equation}
= \bar P_1(s)/[1 - \bar P_1(s)]
\label{2.3}
\end{equation}

Conversely, a given $\rho$ and $g(r)$ will be realizable as a renewal point
process if and only if
\begin{equation}
\bar Q(s) = \rho \bar g(s)/[1 + \rho \bar g(s)]
\label{2.4}
\end{equation}
is the Laplace transform of a probability density, $P_1(r) \geq 0$,
satisfying (\ref{2.1}).  
We will show in the next section that for the one dimensional $g(r)$ given in
(\ref{1.6}) this is true when and only when 
$\rho D \leq e^{-1}$.

It is clear from the definition of a renewal process that the higher
order correlation functions of such a system can be readily expressed
in terms of $\rho$ and $g(r)$.  More specifically given points $x_1 <
x_2 < ... < x_n$ on the line we have for $n = 3,4,...$,
\begin{equation}
\rho_n(x_1,...,x_n)/\rho_{n-1}(x_1,...,x_{n-1}) = \rho g(x_n -
x_{n-1})
\label{2.7}
\end{equation}
since the left side is just the particle density at $x_n$ given that
there are particles at $x_1,...,x_{n-1}$.  Thus
\begin{equation}
\rho_3(x_1,x_2,x_3) = \rho^3 g(x_2-x_1) g(x_3-x_2), \quad x_1 < x_2 <
x_3, 
\label{2.8}
\end{equation}
etc.

There is also a simple expression for $s$, the entropy per unit length
of a renewal process \cite{Cox-Isham},
\cite{Aizenman-Goldstein-Lebowitz}.  It is given by the following
formula, see {Aizenman-Goldstein-Lebowitz}
\begin{equation}
s = -\rho \int_0^\infty P_1(r) \log[P_1(r)/W_0(r)]dr + \rho
\label{2.9}
\end{equation}
where $W_0(r) = \int_r^\infty P_1(y)dy$ is the probability that there
is no particle between $q$ and $(q+r)$.

We can realize an RP as an equilibrium system of particles in $d=1$ in
which only nearest neighbors interact: there are no interactions
between non-nearest neighbor particles, whatever the distances between
them.
{}For such a 
pair interactions $u(r)$, $P_1(r)$ is given by
\cite{Hansen-McDonald}
\begin{equation}
P_1(r) = C e^{-\beta[pr + u(r)]}, \quad r > 0,
\label{2.6}
\end{equation}
where $\beta$ is the reciprocal temperature, $p = p(\beta, \rho)$ is
the pressure and $C = [ \int_0^\infty e^{-\beta[pr +
u(r)]}dr]^{-1}$ is a normalization constant.  Conversely given
$P_1(r)$ we can always define a $\beta u(r)$ and the corresponding
$\beta p$ by inverting (\ref{2.6}).

A well known example of such an equilibrium system with only pair
interactions  is that of hard rods
with diameter $D$, $u(r) = \infty$ for $r < D$, $u(r) = 0$ for $r >
D$.  {}For this system  $P_1(r) = 0$, for $r < D$, and
\begin{equation}
P_1(r) = \beta p e^{-\beta p(r-D)}, \quad {\rm for} \quad r \geq D
\label{2.10}
\end{equation}
with 
\begin{equation}
\beta p = \rho[1 - \rho D]^{-1}.
\label{2.11}
\end{equation}
Eq. (\ref{2.8}) then gives the well known formula for the entropy
density of this system \cite{Hansen-McDonald}
\begin{equation}
s = -\rho \log[\rho/(1 - \rho D)] + \rho
\label{2.12}
\end{equation}

\section{The Realizability of (\ref{1.6}) as an RP}
\setcounter{equation}{0}
By general theorems a necessary and sufficient condition for a
function of $s$, to be the Laplace transform of a non-negative density
is that it be ``completely monotone'' for all $s \geq 0$ \cite{Widder}.  That
is, it is required that its derivatives alternate in sign for all $s
\geq 0$.  Thus for $g$ to define a renewal process it is necessary and
sufficient that $\bar Q(s)$ in (\ref{2.4}) have the property that
\begin{equation}
(-1)^k \bar Q^{(k)}(s) > 0, ~~~~ {\rm for ~~ all} ~~ k =
0,1,2,..., ~~ {\rm and ~~ all} ~~ s \geq 0,
\label{3.1}
\end{equation}
where $f^{(k)}(s) \equiv d^k f(s)/ds^k$.

{}For the $g(r)$ in 
 (\ref{1.6}), 
\begin{equation}
\bar g(s) = \rho \int_D^\infty e^{-sr}dr = \rho s^{-1} e^{-sD}
\label{3.2}
\end{equation}
and the corresponding $\bar Q(s)$ in (\ref{2.4}) is
\begin{equation}
\bar Q(s) = \rho e^{-sD}/[s + \rho e^{-sD}]
\label{3.3}
\end{equation}

It can be shown that (\ref{3.1}) will be satisfied by (\ref{3.3}) if
and only if $\rho D \leq e^{-1}$ \cite{Costin}.  Here we provide a simple
construction of this point process by starting with a Poisson process
on the line, $x \in (-\infty, \infty)$, with density $\lambda$ and
removing points which are too close ending up with a density $\rho =
\lambda e^{- \lambda D}$ and the step $g(r)$ of (\ref{1.6}).

The procedure is as follows.  Denote the points of the Poisson
process, by $...,-x_2, -x_1, x_0, x_1, x_2, x_3, ...$, with $x_i \leq
x_{i+1}$.  Then if \linebreak $(x_{i+1} - x_i) < D$, $x_i$ is removed;
if $(x_{i+1}-x_i) \geq D$, $x_i$ stays.  Now the probability that
$(x_{i+1}-x_i)$ is greater than $D$ is $e^{-\lambda D}$ so the density
of remaining points is
\begin{equation}
\rho D = \lambda D e^{-\lambda D} \leq e^{-1}.
\label{3.4}
\end{equation}
The last inequality follows from the fact that $y e^{-y}$ has its
maximum value $e^{-1}$ at $y=1$.  Note that for $\rho D < e^{-1}$
there are two different values of 
$\lambda$ which lead to the same RP with density $\rho$ (see below).

The new translation invariant process
with density $\rho$ clearly has $g(r) = 0$ for $r < D$.  To see that $g(r) =
1$ for $r > D$ we note that, given a surviving point at position $q$, the
density of other surviving points at $q+r$ is, for $r \geq D$,  just
the density of points for the Poisson processes which have survived,
i.e.\ $\lambda e^{-\lambda D} = \rho$.

It is clear from the above construction that the new process satisfies
the conditions at the beginning of sec.\ 2 and so is an RP with
$P_1(r) = Q(r)$, the inverse Laplace transform of $\bar Q(s)$ in (\ref{3.3}).
To compute $Q(r)$ we use units in which $\rho = 1$. Define
\begin{equation}
Q(r) = Q(y + nD) = w_n(y), \quad {\rm for} ~~ nD \leq r < (n+1)D
\label{3.5}
\end{equation}
and $0 \leq y \leq D$, ~~ $n = 0,1,2,...$.  It is then easy to deduce
from (\ref{3.3}) that 
\begin{equation}
\label{3.6} w_{n+1}(y) = w_{n+1}(0) - \int_0^y w_n(x) dx, \quad n \geq 0
\end{equation}
with
\begin{equation}
w_0(y) = 0, \quad w_1(y) = 1, \quad w_2(y) = 1-y, ...
\label{3.7}
\end{equation}
{}Furthermore
\begin{equation}
w_{n+1}(0) = w_n(D) \quad {\rm for} ~~~ n \geq 1
\label{3.8}
\end{equation}
i.e., $Q(r)$ is continuous for $r > D$.  

Define now,
\begin{equation}
\psi(\lambda;y) = \sum_{n=1}^\infty \lambda^n w_n(y)
\label{3.9}
\end{equation}
It follows then from (\ref{3.6}) that
\begin{equation}
\psi(\lambda,y) = \psi(\lambda;0) - \lambda \int_0^y
       \psi(\lambda;x)dx
\label{3.10}
\end{equation}
and thus
\begin{equation}
\psi(\lambda; y) = \psi(\lambda; 0) e^{-\lambda y}.
\label{3.11}
\end{equation}
Putting $y=D$ then yields
\begin{equation}
\psi(\lambda;D) = \psi(\lambda;0) e^{-\lambda D} =
       \frac{1}{\lambda} [\psi(\lambda;0) - \lambda]
\label{3.12}
\end{equation}
where the last equality follows from (\ref{3.8}) and (\ref{3.7}).  This gives
\begin{equation}
\psi(\lambda;0) = \frac{\lambda}{1 - \lambda e^{-
       \lambda D}}.
\label{3.13}
\end{equation}
The positivity of $Q(r)$ is
equivalent to the requirement that
 all the
coefficients $C_j$ in the expansion of
\begin{equation}
\psi(\lambda;0) = \sum_{j=0}^\infty C_j \lambda^j
\label{3.14}
\end{equation}
are positive.  This again leads to the requirement that $ D \leq
e^{-1}$, with the explicit formula (due to E.\ Speer)
\begin{equation}
w_n(y) = \sum_{k=1}^n [(n-k)D + y]^{k-1} (-1)^{k-1}/(k-1)!, \quad D \leq
e^{-1}.
\label{3.15}
\end{equation}

\section{Determinantal Point Process}
\setcounter{equation}{0} 

We review here briefly how one can obtain point processes from a
$g({\bf r})$ satisfying certain inequalities in any dimension.  The
construction of such processes
is a subject of great current interest in mathematics and we refer the
reader to \cite{Soshnikov} for more information. We again restrict
ourselves to homogeneous systems and choose units in which $\rho=1$.
Let $B({\bf r})$ be a complex function such that
\begin{equation}
B({\bf r}) = B^*(-{\bf r}), \quad B(0) = 1
\label{4.1}
\end{equation}
and
\begin{equation}
0 \leq \hat B(k) \equiv \int_{\RR^d} e^{-i{\bf k}\cdot{\bf r}}
B(r)d{\bf r} \leq 1
\label{4.2}
\end{equation}
It can then be proven that conditions (\ref{4.1}) and (\ref{4.2}) are
necessary and sufficient for the existence of a point process with
$n$-particle densities given by the determinants \cite{Soshnikov}
\begin{equation}
\rho_n({\bf r}_1,...,{\bf r}_n) = \left |\begin{array}{cccc}
1 & B({\bf r}_{12}) & ... & B({\bf r}_{1n}) \cr
B({\bf r}_{21}) & 1 & ... & B({\bf r}_{2n}) \cr
~~~ & ....... & ~~~ \cr
B({\bf r}_{n1}) & B({\bf r}_{n2}) & ... & 1 \cr
\end{array}\right|.
\label{4.3}
\end{equation}
where ${\bf r}_{ij} = {\bf r}_i - {\bf r}_j$.  In particular we have
\begin{equation}
g({\bf r}) = 1 - |B({\bf r})|^2
\label{4.4}
\end{equation}
with $g(0) = 0$, $g(r) \leq 1$.

Such a process is called a determinantal point process (DP). 
It follows that given 
a $g({\bf r})$, such that the Fourier
transform of $B({\bf r}) \equiv [1 - g(r)]^{1/2}$ satisfies
(\ref{4.2})
and $g(0) = 0$, $0 \leq g(r) \leq 1$, 
 we can construct 
 a point process with
explicit correlations (\ref{4.3}).  This gives a large (uncountable) class of
$g({\bf r})$ which have the realizability property. {}For all
details, see \cite{Soshnikov} and references there.

We make two remarks:

1) To obtain a superhomogeneous system
    \cite{Lebowitz}--\cite{Aizenman-Goldstein-Lebowitz} with $\hat
    S(0) = 0$, for the determinantal point process specified by some
    $B({\bf r})$, it is necessary and sufficient that $\hat B({\bf
    k})$ be a characteristic function of a set $\Omega$ in ${\bf
    k}$-space, i.e.\ $\hat B({\bf k}) = 1$ for ${\bf k} \in \Omega$,
    $\hat B({\bf k}) = 0$ otherwise.  This is the case for the well
    known one dimensional system of particles on a circle with pair
    interaction $\phi(r_{ij}) = -e^2 \log|r_{ij}|$, at reciprocal
    temperature $\beta = 2e^{-2}$.  {}For this system, with $\rho=1$,
    the infinite volume limit of the radial distribution function is
    given by $g(r) = 1 - (\sin \pi r/\pi r)^2$ and the variance $V_L$
    of the number of particles in an interval of length $L$ grows like
    $\log L$.  This system is sometimes referred to as the Dyson gas:
    the $\rho_n$ describe the correlations of the eigenvalues of
    random Gaussian Hermitian Matrices \cite{Mehta},
    \cite{Costin-Lebowitz}.

2.  To get translation invariant determinantal correlation functions
    as in (\ref{4.3}) it is not necessary that $B$ depend only on
    ${\bf r}_{12}$.  It is only necessary that $B({\bf r}_1, {\bf
    r}_2) = F({\bf r}_{12}) e^{i[\phi({\bf r}_1)-\phi({\bf r}_2)]}$
    with $B({\bf r}_1,{\bf r}_2)$ satisfying, as an operator, the
    analogue of (\ref{4.1}) \cite{Soshnikov},
    \cite{Soshnikov-private}.  This is the case for a two dimensional
    one component plasma with $\phi(r_{ij}) = -e^2 \log|r_{ij}|, ~~
    \beta = 2e^{-2}$ \cite{Lebowitz}, \cite{Mehta}.  {}For this system
    the variance in the number of particles in a disc of radius $R$
    grows like $R$ \cite{Lebowitz}.

\section{Example and Discussion}
\setcounter{equation}{0} 

We illustrate here the construction of a DP in $d$ dimensions from a
given $\rho_1$ and $\rho_2$ which is, in $d=1$, also an RP.  As in the
example (\ref{1.6}) this can be done for only a subset of the
parameter for which (\ref{1.1}), (\ref{1.2}) and (\ref{1.4}) are
satisfied.  On the other hand everything here can be computed
explicitly in an elementary way.  Using units in which $\rho = 1$, let
\begin{equation}
g(r) = 1 - e^{-\lambda r}, \quad \lambda \geq 0.
\label{5.1}
\end{equation}
It is easily checked that this $g$ satisfies (\ref{1.1}), 
(\ref{1.2}) and (\ref{1.4}), whenever $\lambda \geq \lambda_d$,
$\lambda_1 = 2, \lambda_3 = (8 \pi)^{1/3},...$.  It follows from
(\ref{4.4}) that this $g$ determines a DP with $B = e^{-\lambda r/2}$
whenever $\lambda \geq 2 \lambda_d$.

On the other hand, using
(\ref{2.4}), we get for $d=1$, the Laplace transform, 
\begin{equation}
\bar Q(s) = \lambda/[s^2 + \lambda s + \lambda]
\label{5.2}
\end{equation}
from which one readily finds, by factorizing the denominator in
(\ref{5.2}) and using criteria (\ref{3.1}), that (\ref{5.1})
determines a RP if and only if $\lambda \geq 2 \lambda_1 = 4$.  {}For
such values of $\lambda$, it is then easily found that

\begin{subeqnarray}\label{5.3a}
P_1(r) =\lambda[\lambda^2 - 4\lambda]^{-1/2} e^{-\lambda r} \Big
\{\exp[\frac{-\lambda + \sqrt{\lambda^2 - 4\lambda}}{2}r]\nonumber\\ -
\exp [\frac{-\lambda - \sqrt{\lambda^2 - 4\lambda}}{2}r]\Big \}. \quad
\quad \lambda > 4,\\ P_1(r) = 4r e^{-2r}. \quad \lambda = 4
\label{5.3b}
\end{subeqnarray}

It is now 
easy to check that (\ref{2.7}) and (\ref{4.3}) give the same
correlations $\rho_n(x_1,...,x_n)$.  In particular 
for all $\lambda \geq 4$,
\begin{equation}
\rho_3(x_1,x_2,x_3) = (1 - e^{-\lambda(x_2-x_1)})(1 -
e^{-\lambda(x_3-x_2)}), ~~ x_1 \leq x_2 \leq x_3
\label{5.4}
\end{equation}
Using (\ref{5.3a}) in (\ref{2.9}) one can also obtain the entropy of
this system for $\lambda \geq 4$.  

The fact that the DP and RP constructions in the
above example yield the same point process might suggest that 
an RP or DP determines a unique point process.
This may indeed be the case.  We note, however, that uniqueness is not
true in general, as can be seen from considerations of systems with
non-reflection invariant correlations.  Thus, while we always have, for
translation invariant systems, that $g({\bf r}) = g(-{\bf r})$, there
is no such symmetry for the higher order $\rho_n$.  In particular  there is
an explicit construction of a translation invariant point processes in
$d=1$ which, when run ``backwards'' will have the same $\rho$ and
$g(r)$ as the original process but a $\rho_3^\prime$ obtained from the
original one by reflection, i.e.\ $\rho^\prime_3(x_1,x_2,x_3) =
\rho_3(-x_1,-x_2,-x_3) \ne \rho_3(x_1,x_2,x_3)$; see
\cite{Goldstein-Lebowitz-Speer} for 
details.  

The sufficiency of conditions (\ref{1.1}), (\ref{1.2}) and (\ref{1.4})
(with the generalizations given in the Appendix) remains open although
it seems unlikely that any finite number of conditions would suffice
for the general case \cite{Percus}, \cite{Yamada}, 
\cite{Kuna}.  The construction of the step $g(r)$ in (\ref{1.6}) by
the dilution of a Poisson process does not seem to work in $d > 1$.
On the other hand we have not found any counterexample so far.

\bigskip \bigskip
\noindent {\it Acknowledgments}: We benefited greatly from
discussions with S.\ Goldstein, T.\ Kuna, J.\ Percus, E.\ Speer, F.\
Stillinger and S.\ Torquato.  We also thank A. Soshnikov for
enlightening us about DP. Work supported by NSF Grant DMS-0100495,
DMS-0074924, and DMR 01-279-26, and AFOSR Grant AF 49620-01-1-0154.

\noindent {\bf Appendix:  Proof of (\ref{1.4})}

We give here an elementary proof of (\ref{1.4}) and its
generalizations, see also \cite{Yamada}.  
Let $P(k)$ = Prob. of having $k$ particles in $\Lambda$ such that
$$
\langle k \rangle = \sum_{k=0}^\infty kP(k) = N + \theta, \quad
N = 0,1,2,..., ~~ 0 \leq \theta \leq 1.
$$
Then the variance
$$ 
V_\Lambda = \sum_k [k - N - \theta]^2 P(k)
$$
$$
~~ = \sum (k - N)^2 P(k) - 2 \theta[\langle k \rangle - N] + \theta^2
$$
$$
~~ = \sum (k - N)^2 P(k) - \theta^2
$$
$$
~~ \geq \sum (k - N) P(k) - \theta^2 = \theta(1 - \theta). 
$$
The inequality follows from the fact that $n^2 \geq n$ for $n$ an integer.
Equality occurs when $P(k) = \alpha \delta_{k,N} + (1 -
\alpha)\delta_{k,N+1}$, with $\alpha$ determined by $\theta$.

We note that the same argument works also for the variance of a linear
combination of the number of particles $N_{\Lambda_i}$ in 
regions $\Lambda_i$.  Let $Y = \sum_{i=1}^k m_i N_{\Lambda_i}$, where $m_i$
are integer coefficients, then $\langle(Y - \langle Y \rangle )^2
\rangle \geq \theta(1 - \theta)$.  In particular consider the difference
between the number of particles in a region $\Lambda_1$ and a region
$\Lambda_2$.  Letting $\langle (N_{\Lambda_1} - N_{\Lambda_2}) \rangle
= K + \theta$, ~~~ $K = 0, \pm 1, \pm 2, ...$ we again have
$$
V_{1,2} = \langle [N_{\Lambda_1} - N_{\Lambda_2} - K - \theta]^2
\rangle \geq \theta(1 - \theta)
$$

\end{document}